\documentclass[12pt,letterpaper]{article}
\usepackage[a4paper, total={7in, 10in}]{geometry}

\usepackage{graphicx}
\usepackage{helvet}
\usepackage{authblk}
\usepackage{hyperref}
\usepackage{amsmath} 
\usepackage{amssymb} 
\usepackage{orcidlink} 
\usepackage[super,comma,sort&compress] 
 {natbib}
\usepackage[right]{lineno} \linenumbers

\usepackage[T1]{fontenc} 
\usepackage{url}   
\usepackage{booktabs}  
\usepackage{amsfonts}  
\usepackage{nicefrac}  
\usepackage{microtype}  
\usepackage{xcolor}   
\usepackage{multirow}
\usepackage{threeparttable}
\usepackage{amsmath}
\usepackage{subcaption}
\usepackage{makecell}
\usepackage{placeins}
\usepackage{listings}
\usepackage{xcolor}

\usepackage{makecell}

\makeatletter
\renewcommand{\maketitle}{\bgroup\setlength{\parindent}{0pt}
\begin{flushleft}
 \textbf{\@title}
 
 \@author
\end{flushleft}\egroup}
\makeatother

\title{MEETI: A Multimodal ECG Dataset from MIMIC-IV-ECG with Signals, Images, Features and Interpretations}

\date{}

\author[1,\#\orcidlink{0000-0002-4041-3083}]{Deyun Zhang}
\author[2,\#]{Xiang Lan}
\author[1]{Shijia Geng}
\author[3]{Qinghao Zhao}
\author[4]{Sumei Fan}
\author[2,*]{Mengling Feng}
\author[5,6,*,\orcidlink{0000-0001-7521-5127}]{Shenda Hong}

\affil[1]{HeartVoice Medical Technology, Hefei, China}
\affil[2]{Saw Swee Hock School of Public Health and Institute of Data Science, National University of Singapore, Singapore}
\affil[3]{Department of Cardiology, Peking University People’s Hospital, Beijing, China}
\affil[4]{College of Integrative Chinese and Western Medicine, Anhui University of Chinese Medicine, Hefei, China}
\affil[5]{National Institute of Health Data Science, Peking University, Beijing, China}
\affil[6]{Institute for Artificial Intelligence, Peking University, Beijing, China. }

\affil[$\#$]{These authors contributed equally}
\affil[*]{Correspondence: hongshenda@pku.edu.cn, ephfm@nus.edu.sg}

\begin{document}

\maketitle

\section*{Abstract}

Electrocardiogram (ECG) plays a foundational role in modern cardiovascular care, enabling non-invasive diagnosis of arrhythmias, myocardial ischemia, and conduction disorders. While machine learning has achieved expert-level performance in ECG interpretation, the development of clinically deployable multimodal AI systems remains constrained, primarily due to the lack of publicly available datasets that simultaneously incorporate raw signals, diagnostic images, and interpretation text. Most existing ECG datasets provide only single-modality data (e.g., raw waveform signals) or, at most, dual modalities (e.g., image-text pairs), making it difficult to build models that can understand and integrate diverse ECG information in real-world settings. To address this gap, we introduce MEETI (MIMIC-IV-Ext ECG-Text-Image), the first large-scale ECG dataset that synchronizes raw waveform data, high-resolution plotted images, and detailed textual interpretations generated by large language models (LLMs). In addition, MEETI includes beat-level quantitative ECG parameters extracted from each lead, offering structured parameters that support fine-grained analysis and model interpretability. MEETI is built upon MIMIC-IV-ECG, one of the largest open-access collections of 12-lead clinical ECG recordings, comprising over 800,000 ten-second recordings paired with expert reports. MEETI expands this foundation with three key additions: high-resolution ECG images, beat-level quantitative ECG parameters, and detailed textual interpretations. Each MEETI record is aligned across four components: (1) the raw ECG waveform, (2) the corresponding plotted image, (3) extracted feature parameters, and (4) detailed interpretation text. This alignment is achieved using consistent, unique identifiers. This unified structure supports transformer-based multimodal learning and supports fine-grained, interpretable reasoning about cardiac health. By bridging the gap between traditional signal analysis, image-based interpretation, and language-driven understanding, MEETI established a robust foundation for the next generation of explainable, multimodal cardiovascular AI. It offers the research community a comprehensive benchmark for developing and evaluating ECG-based AI systems.


\section*{Background \& Summary}

Cardiovascular diseases (CVDs) are the leading cause of mortality worldwide \cite{martin20252025}, contributing to over 17 million deaths each year. The electrocardiogram (ECG) remains the primary non-invasive modality to assess cardiac electrophysiology \cite{tan2025pediatric}. It provides temporal waveform information that is critical for diagnosing arrhythmias, myocardial ischemia, and conduction abnormalities \cite{yagi2024routine, zheng202012}. However, manually interpreting large volumes of ECG recordings is time-consuming and prone to inter-observer variability. Recently, machine learning and artificial intelligence (AI) approaches have emerged as powerful tools for ECG analysis \cite{hong2020opportunities,lan2022intra,lan2024towards,li2025electrocardiogram}. They can achieve expert-level performance in waveform analysis, arrhythmia detection, and risk stratification. These advances underscore the urgent need for comprehensive, multimodal ECG datasets. Such datasets are essential to ensure reproducible research and to accelerate clinical translation. Over the past few years, multimodal large language models (LLMs) have demonstrated remarkable performance in interpreting complex medical imaging data\cite{liu2023medical}. They can seamlessly integrate radiographic, histopathological, and clinical text inputs to support diagnostic and prognostic workflows \cite{sandmann2025benchmark, zhou2025large}. These advanced architectures use transformer-based encoders to fuse pixel-level features with semantic context. As a result, tasks such as lesion detection, disease classification, and report generation can be performed with near-human precision\cite{hong2025evaluating}. Despite these successes, current LLMs remain incapable of directly processing and understanding electrocardiographic waveforms. Cardiac signals have unique temporal dynamics and amplitude variations that differ fundamentally from spatial image patterns. Bridging this gap requires specialized datasets and model designs that jointly represent ECG signals alongside complementary textual and visual information.

Most existing ECG datasets provide only single-modality data (e.g., raw waveform signals) or, at most, dual modalities such as image–text pairs, as commonly seen in clinical reports. However, contemporary AI research increasingly requires multimodal representations to model the complex nature of ECG data, which inherently combine temporal waveforms, visual patterns, and diagnostic text \cite{wagner2020ptb, gow2023mimic, kansal2025mc}. This mismatch between available data formats and modeling needs has hindered progress in developing explainable and generalizable ECG-based AI systems. Models trained on digitized time series cannot easily leverage the many image‐based clinical interpretations available online \cite{al2023machine, chen2024congenital}. Traditional analysis pipelines rely on handcrafted parameter extraction (e.g., QRS complex duration, ST segment deviations) to support diagnostic decisions\cite{fan2025detecting}. In contrast, deep learning models act as opaque black boxes without explicit feature semantics\cite{csahin2025unlocking}. These two limitations highlight that ECG data naturally combine visual, temporal, and textual information. As a result, we need a unified multimodal approach. Training multimodal language models to understand ECG data requires genuine ECG multimodal examples—paired waveform images, raw signals, and textual annotations \cite{li2024frozen, lai2025diffusets}. However, existing public ECG datasets only include signal–text pairs and leave out image data \cite{tsutsui2025shdb, zheng2020121}. This gap prevents a systematic study of how to combine visual and temporal cardiac representations in transformer-based models. The lack of integrated datasets is therefore a bottleneck for ECG multimodal research. It motivates the creation of resources that encode images, signals, and semantics together for end-to-end model training.

To address these limitations, we introduce MEETI (MIMIC-IV-Ext ECG-Text-Image, Figure 1), a multimodal ECG dataset that enables LLMs to process and interpret ECG data more effectively. MEETI is built on the MIMIC-IV-ECG dataset\cite{gow2023mimic, johnson2020mimic}, one of the largest and most representative public 12-lead clinical ECG collections available. It covers a wide range of cardiac conditions and signal qualities that reflect real-world clinical settings. MEETI is the first multimodal ECG dataset to provide synchronized access to four key components: raw ECG signals, plotted waveform images, extracted quantitative parameters, and detailed interpretation texts. By aligning these modalities, MEETI establishes a unified framework that integrates signal, visual, and language representations. This framework standardizes input formats for training LLMs and promotes explainable AI by enabling clinically interpretable reasoning, rather than replying on traditional `black-box' classification approaches. The integration of visual, temporal, and semantic modalities facilitates cross-modal alignment and supports the development of models capable of performing ECG analysis from multiple perspectives. Moreover, MEETI encourages a paradigm shift from conventional end-to-end diagnostic pipelines toward inference-based reasoning. Ultimately, this resource enhances the ability of multimodal LLMs to generate meaningful, context-aware interpretations of cardiac health.

\begin{figure}
 \includegraphics[width=1.0\linewidth]{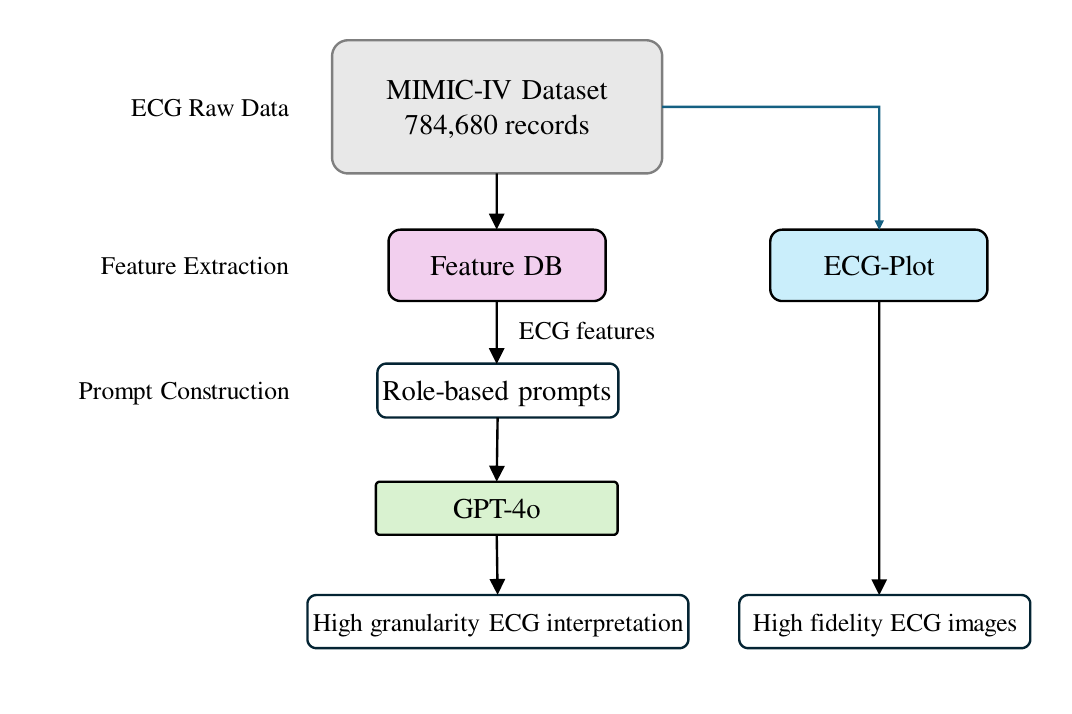}
 \caption{Schematic overview of the components that constitute the MEETI dataset.}
 \label{fig1}
\end{figure}

\section*{Methods}

\subsection*{ECG Signals}

As part of routine clinical care at the Beth Israel Deaconess Medical Center (BIDMC), diagnostic 12-lead ECGs were collected using standard ECG machines and subsequently archived in the MIMIC-IV-ECG\cite{gow2023mimic}. This dataset comprises approximately 800,000 ten-second, 12-lead ECG recordings sampled at 500 Hz, matched to nearly 160,000 unique patients included in the broader MIMIC-IV Clinical Database. In addition to the raw waveform data, machine-generated measurements and cardiologist-interpreted reports are also available for a substantial proportion of ECGs. All records have been de-identified in accordance with the HIPAA Safe Harbor requirements to ensure patient privacy.

\subsection*{ECG Parameter Extraction}

To extract a comprehensive set of quantitative descriptors from the MIMIC-IV-ECG waveforms, we integrate the open source FeatureDB toolkit (\url{https://github.com/PKUDigitalHealth/FeatureDB}) into our pre-processing pipeline \cite{fan2025detecting, hong2020cardiolearn}. FeatureDB employs a multi-stage workflow beginning with adaptive peak detection to localize the P wave, QRS complex, and T wave across each lead. Following initial peak localization, discrete wavelet transforms are applied at multiple scales to refine the onset and offset boundaries of these waveform components, thereby ensuring precise measurement of key temporal intervals (PR interval, QRS duration, QT interval, and QTc). We applied this pipeline to all de-identified 12 lead ECG records after verifying signal integrity (i.e., complete lead set, consistent sampling rate, and absence of major artifacts). Extracted parameters were stored in a structured relational database, with each record linked to its raw WFDB waveform files and corresponding expert reports via unique study identifiers. This systematic application of FeatureDB not only facilitates scalable, reproducible extraction of clinically meaningful ECG metrics but also lays the groundwork for downstream multimodal integration with image and text modalities, ultimately supporting the development of interpretable AI models in cardiac electrophysiology.

\subsection*{ECG Plot}

To generate high-fidelity ECG images, we rendered each preprocessed 12 lead waveform into a standardized graphical format using the open‑source ecg plot Python library (\url{https://github.com/PKUDigital Health/ecg_plot}) \cite{zhang2023artificial}. For each record, we configured the plotting parameters to emulate clinical paper recordings: a paper speed of 25 mm/s and an amplitude scaling of 10 mm/mV. Under these settings, each large grid block corresponds to 0.5 mV in voltage and 0.2 s in time, thereby ensuring precise spatial encoding of waveform morphology. Leads were arranged in the conventional $3\times4$, $6\times2$, or $12\times1$ (I, II, III; aVR, aVL, aVF; V1-V6) with appropriate inter‑panel spacing and grid visibility. Images were exported at 300 dpi resolution to facilitate downstream vision‑based model training and to maintain pixel-level detail of low-amplitude features. Each image file was named using the study identifier, then stored alongside the corresponding waveform and text report for seamless multimodal alignment. This procedural standardization yields a uniform image corpus suitable for transformer-based and convolutional model development in ECG interpretation.

\subsection*{ECG Interpretation Generation}

Human-written ECG reports are typically brief and often lack the detail needed for comprehensive interpretation.
To produce high-granularity ECG interpretations that explicitly link diagnoses to measurable parameters such as QRS durations or PR intervals, we develop role-based prompts that, combined with extracted ECG parameters, allow GPT-4o \cite{achiam2023gpt} to effectively utilize its latent medical knowledge for generating clinically detailed and parameter-grounded interpretations.
For each ECG record, the extracted parameters are structured into sequential data. For example, an ECG signal containing ten visible heartbeats allows us to construct a QRS duration sequence as \([QRS_1, QRS_2, ..., QRS_{10}]\), where each element precisely represents the QRS duration for a individual heartbeat. These sequences offer fine-grained descriptions of ECG physiology, enabling assessment of cardiac function at the individual heartbeat level.
The role-based prompts then guide GPT-4o \cite{achiam2023gpt} with cardiology-specific instructions to analyze various aspects of the ECG parameters, conditioned on the ground-truth report. For example, the prompt may instruct GPT-4o \cite{achiam2023gpt} to examine P wave amplitude and duration in Lead II for signs of atrial enlargement. These instructions also reflect the logical structure of clinical diagnostic workflows.
Through this approach, we are able to generate high-granularity ECG interpretations at scale without relying on manual annotations.

\section*{Data Records}

\subsection*{Data Released as Part of This Dataset}

MEETI can be accessed via the following link: \url{https://github.com/PKUDigitalHealth/MIMIC-IV-ECG-Ext-Text-Image}. MEETI consists of four complementary components: raw ECG signals and report, plotted ECG images, per-beat ECG parameters, and LLM-based textual interpretations. Each component is linked by a unique record identifier so that users can cross-reference signals, images, parameters, and interpretations. Our goal is to lower the barrier to entry for multimodal ECG research. By making all four data types available in one cohesive resource, we hope to accelerate developments in ECG-based diagnostics, prognostics, and educational tools. Researchers can seamlessly integrate waveform analysis, image-based deep learning, feature engineering, and natural language processing within a single dataset. MEETI thus represents a comprehensive, publicly accessible resource for advancing AI in cardiology.

\begin{itemize}
\item \textbf{Raw ECG data}: The raw ECG signals and accompanying text reports are included directly from MIMIC-IV-ECG. These original signals cover a wide range of patient demographics, clinical conditions, and recording settings. Researchers can access the same high-resolution waveform data that clinicians used at Beth Israel Deaconess Medical Center. By depositing MEETI on PhysioNet, we ensure long-term accessibility and enable others to reproduce our results without additional data collection efforts.

\item \textbf{Plotted ECG image}: MEETI provides a set of plotted ECG images. Each image arranges twelve leads in a $12\times1$ grid, replicating the standard clinical layout. We generated these plots using the matplotlib library. The plotting scripts apply consistent scaling and labeling conventions to preserve the clinical context. Clinicians and machine learning practitioners alike can use these images for visual analysis, teaching, or training convolutional neural networks. Since plotted images are derived directly from the same high-fidelity signals, they faithfully represent the original ECG morphology.

\item \textbf{ECG parameter}: MEETI also includes quantitative ECG parameters for every beat in each lead. These parameters were automatically extracted using the FeatureDB toolkit. For each lead, FeatureDB identifies fiducial points (P-wave, QRS complex, T-wave) and computes intervals (PR, QRS duration, QT/QTc) as well as amplitudes (R- and S-wave heights in leads V1 and V5, for example). By capturing beat-by-beat variability, these detailed metrics support both statistical analyses and deep learning approaches. Researchers interested in heart rate variability, conduction delays, or repolarization abnormalities will find this dataset especially useful. Providing per-beat parameters from all twelve leads means that MEETI can facilitate studies of subtle waveform changes that might be missed when relying on summary statistics alone.

\item \textbf{Interpretation of the LLM}: MEETI offers natural language interpretations generated by an LLM. We used GPT-4o \cite{achiam2023gpt} to produce these interpretations based on the combined input of the original textual reports and the extracted ECG parameters. For each recording, the LLM output highlights significant findings, such as rhythm abnormalities or evidence of ischemia. By comparing LLM-generated text with clinician-authored reports, users can evaluate the model’s diagnostic suggestions. These interpretations serve as a benchmark for future work on AI-driven ECG reporting, and they demonstrate how multimodal inputs (e.g., text plus structured parameters) can inform more accurate, context-aware explanations.
\end{itemize}

Each subject’s data is organized into a dedicated folder named by their unique subject ID, which corresponds directly to the IDs used in the MIMIC-IV-ECG database. Within each folder, plotted ECG images are saved as individual PNG files, with filenames that reflect lead arrangement and recording timestamps. Beat-by-beat parameters—such as PR interval, QRS duration, QT/QTc values, and fiducial point amplitudes—are stored in the MAT file. This file includes detailed metrics for every beat across all twelve leads, making it easy to perform quantitative analyses without further preprocessing. The LLM-generated textual interpretations from GPT-4o \cite{achiam2023gpt} are also provided in the MAT file. Those interpretations link directly to the corresponding ECG parameter sets, allowing users to compare AI-driven commentary with numerical results. By grouping all PNG and MAT files under a single subject-specific directory (Table 1), MEETI ensures that researchers can quickly locate ECG images, per-beat ECG parameters, and LLM-based textual interpretations for each individual  (Table 2). This streamlined structure supports seamless multimodal analysis.

\subsection*{Descriptive Statistics}

Except for a small number of samples that could not be processed by the algorithm, MEETI cover the full MIMIC-IV-ECG dataset, up to 784,680 records from 160,597 patients.

We used several tables and figures to present the ECG parameters included in MEETI. Table 3 shows the median values of all ECG parameters for each complete ECG recording. These statistics provide a clear overview of the central tendencies across subjects and recordings. Researchers interested in overall waveform characteristics can refer to Table 3, while those focused on lead-specific details will find Table 3 especially useful. All of these parameters are stored in the .mat files for easy access. The clear linkage between the tables, figures, and .mat files ensures that datasets are both transparent and reproducible (Table 3, Figure 2).

The ECGs are grouped into subdirectories based on subject\_id. Each record path follows the pattern: files/pNNNN/pXXXXXXXX/sZZZZZZZZ/ZZZZZZZZ, where:

NNNN is the first four characters of the subject\_id,
XXXXXXXX is the subject\_id,
ZZZZZZZZ is the file\_name

An example of the file structure is as follows:

\begin{lstlisting}[language=Python]
 MEETI
 |---p1000
     |---p10000032
         |---s40689238
             |---40689238.mat
 |......
 |---p1571
     |---p15718729
         |---s40000369
             |---40000369.mat
             |---40000369.png
         |---s40461152
             |---40461152.mat
             |---s44659160
         |---44659160.mat
             |---s48122836
             |---48122836.mat
\end{lstlisting}

Above we find two subjects p10000032 (under the p1000 group level directory) and p15718729 (under the p1571 group level directory). For subject p10000032 we find one study: s40689238. For p15718729 we find four studies: s40000369, s40461152, s44659160, s48122836. The study identifiers are completely random, and their order has no implications for the chronological order of the actual studies. Each study has a like-named .png and .mat file (Table 1, Table 2).

\begin{table}[!htbp]
\centering
\caption{Description of key values of the provided datasets.}
\label{tab1}
\renewcommand{\arraystretch}{2}
{%
\begin{tabular}{c c c}
\toprule
 & key value & Description \\
\midrule
.png file & --- & ECG image generated from ECG signal data \\
\midrule
\multirow{4}{*}{.mat file} & id & MIMIC-IV-ECG identifier \\
\cline{2-3}
 & LLM\_Interpretation & \makecell{ECG interpretation generated by LLM based \\ on input prompts} \\
\cline{2-3}
 & report & ECG report from MIMIC-IV-ECG \\
\cline{2-3}
 & featuredb\_lead\_X & \makecell{Beat parameters were extracted from each lead using \\ FeatureDB. 'X' denotes the lead name, which can \\ be one of the following: I, II, III, aVR, aVL, aVF, V1, \\ V2, V3, V4, V5, or V6} \\
\bottomrule
\end{tabular}
}
\end{table}

\begin{table}[h]
\centering
\caption{Descriptions of ECG parameters were extracted from each lead using FeatureDB.}
\label{tab2}
\renewcommand{\arraystretch}{1.5}
{%
\begin{tabular}{c c}
\toprule
ECG parameter name & Description \\
\midrule
HR & \makecell{The heart rate was calculated based on the average of RR1 and RR2, \\ along with the sampling rate} \\
\midrule
RR1 & \makecell{The RR interval between the current heartbeat and the previous one} \\
\midrule
RR2 & \makecell{The RR interval between the current heartbeat and the next one} \\
\midrule
P\_amplitude & \makecell{The amplitude of the P wave, measured from the isoelectric baseline \\ to its peak} \\
\midrule
P\_duration & The duration of the P wave, from onset to its end \\
\midrule
PR\_interval & The interval from the start of the P wave to the start of the QRS complex \\
\midrule
QRS\_amplitude & The peak amplitude of the QRS complex \\
\midrule
QRS\_duration & The duration of the QRS complex, from its onset to its end \\
\midrule
T\_amplitude & The amplitude of the T wave was measured from baseline to peak \\
\midrule
T\_duration & The duration of the T wave, from onset to its end \\
\midrule
ST\_duration & \makecell{The length of the ST segment, from the end of the QRS complex to \\ the beginning of the T wave} \\
\midrule
ST\_form & \makecell{The morphological pattern of the ST segment (e.g., horizontal, \\ upsloping, downsloping)} \\
\midrule
QT\_interval & The interval from the start of the QRS complex to the end of the T wave \\
\midrule
QTc & The QT interval was corrected for heart rate \\
\bottomrule
\end{tabular}
}
\end{table}

The record\_list.csv file contains the file name and path for each record. It also provides the corresponding subject\_id and file\_name. The subject\_id can be used to link a subject from MEETI to the other modules in the MIMIC-IV and MIMIC-IV-ECG Database.

Below is a basic script for reading the downloaded records from this project:

\begin{lstlisting}[language=Python]
 from PIL import Image
 from scipy.io import loadmat
 
 loaddata = loadmat('./p1571/p15718729/s40000369/40000369.mat')
 print(loaddata.keys())
 print(loaddata['id'])
 print(loaddata['LLM_Interpretation'])
 print(loaddata['report'])

 img = Image.open('./p1571/p15718729/s40000369/40000369.png')
 print(img.size)
\end{lstlisting}


\section*{Technical Validation}

The technical validation of MEETI includes two main components. First, we present the distribution and descriptive statistics of the extracted ECG parameters. For each ECG parameter, such as PR interval, QRS duration, QT interval, and T-wave axis, we calculate the mean and standard deviation. Distributions are visualized across all twelve leads to show consistency and variability within the dataset. This step helps ensure that the extracted parameters fall within expected clinical ranges and reflect real-world physiological diversity. Second, we demonstrate the richness of MEETI by showcasing examples of the dataset’s four core components: ECG image, ECG report, beat-level parameters, and LLM interpretations. We highlight how these elements are linked through shared identifiers and are consistently formatted for ease of use. Together, they provide a comprehensive view of each cardiac event, from waveform morphology to diagnostic explanation. This combination supports multiple downstream tasks, such as signal analysis, image classification, and natural language modeling. By validating both the statistical integrity and multimodal structure of the data, we aim to show that MEETI is a reliable and versatile resource for training and evaluating ECG-based machine learning models.

\textbf{ECG parameters: distribution of extracted parameters.} To validate the quality of the extracted ECG parameters, we analyzed their statistical distributions across the entire MEETI dataset. For each parameter—such as PR interval, QRS duration, QT interval, QTc, heart rate, etc.—we calculated the mean, standard deviation, median, interquartile range, etc. These statistics were computed at both the recording level. We also report the missing data ratio, which was effectively zero across all records, indicating high data completeness. Most parameters showed distributions consistent with known physiological ranges. For example, the median PR interval was approximately 140 ms, and the QRS duration clustered around 90 ms, as expected in standard adult ECGs.

These results are presented in Table 3, providing an overview of the central tendency and variability across the entire dataset. The clear, clinically plausible patterns observed across metrics confirm that the extracted parameters are reliable and suitable for downstream modeling tasks such as classification, clustering, or trend analysis in machine learning applications.

\begin{table}[!htbp]
 \centering
 \caption{Statistical analysis of electrocardiogram parameters was conducted by extracting data from MIMIC-IV-ECG using FeatureDB.}
 \label{tab3}
 \renewcommand{\arraystretch}{1.5}
 \resizebox{\textwidth}{!}{ 
 \begin{tabular}{c c c c c c c c c}
 \hline
  ~ & Mean & Std & 95\%CI & Median & IQR & Skewness & Kurtosis & Missing ratio \\ \hline
  HR & 78.19 & 19.77 & 78.17-78.20 & 74.50 & 24.00 & 1.13 & 1.78 & 0.00 \\ 
  RR1 & 403.44 & 94.38 & 403.38-403.50 & 400.00 & 128.50 & 0.38 & 1.17 & 0.00 \\
  RR2 & 403.70 & 94.41 & 403.64-403.76 & 400.5 & 128.00 & 0.41 & 1.72 & 0.00 \\
  P\_amplitude & 0.30 & 0.43 & 0.30-0.30 & 0.10 & 0.31 & 2.95 & 12.33 & 0.00 \\
  P\_duration & 94.01 & 9.42 & 94.01-94.02 & 94.00 & 12.00 & -0.18 & -0.16 & 0.00 \\
  PR\_interval & 135.07 & 60.76 & 135.03-135.11 & 140.00 & 61.00 & -0.58 & 0.83 & 0.00 \\ 
  QRS\_amplitude & 0.59 & 0.62 & 0.59-0.59 & 0.49 & 0.96 & 1.21 & 2.32 & 0.00 \\
  QRS\_duration & 84.65 & 36.48 & 84.63-84.68 & 91.00 & 70.00 & 0.06 & -1.20 & 0.00 \\
  T\_amplitude & 0.34 & 0.42 & 0.34-0.34 & 0.20 & 0.32 & 3.54 & 20.94 & 0.00 \\
  T\_duration & 145.15 & 18.45 & 145.14-145.16 & 144.00 & 26.00 & -0.03 & 0.74 & 0.00 \\
  ST\_duration & 133.10 & 71.08 & 133.06-133.15 & 127.00 & 75.00 & 0.98 & 4.33 & 0.00 \\ 
  ST\_form & 0.00 & 1.00 & 0.00-0.00 & 0.00 & 1 .00 & -0.58 & -1.00 & 0.00 \\
  QT\_interval & 363.98 & 85.51 & 363.92-364.03 & 371.00 & 94 .00 & -0.09 & 2.04 & 0.00 \\ 
  QTc & 409.30 & 91.74 & 409.24-409.35 & 418.00 & 94.5 0 & -0.27 & 0.76 & 0.00 \\
 \hline
 \end{tabular}}
\end{table}

\textbf{Multimodal components: Synchronized ECG signals, waveform images, beat-level quantitative parameters, and interpretation texts generated by LLMs.} We confirmed that each MEETI record reliably links four data modalities around a single visual workspace. In Figure 2, the twelve‑lead ECG image appears on the left. To its right, we display the corresponding per-beat parameters and the clinician’s original text report side by side. This arrangement allows users to see numerical metrics—such as PR interval, QRS duration, and wave amplitudes—aligned directly with expert observations. Beneath the entire panel, we place the GPT-4o generated interpretation. This parameter-based summary references specific parameter values and mirrors the clinical report’s structure, for example: ``The rhythm suggests atrial fibrillation, characterized by the irregular RR intervals observed throughout the ECG. This arrhythmia is further supported by the absence of distinct P waves, vals observed throughout the ECG'' or ``Nonspecific ST-T wave changes are present, particularly noticeable in the lateral leads (V5-V6), which display a subtle upslope in the ST segment, although without a clear pattern of ischemia or infarction.''

\begin{figure}[!ht]
 \includegraphics[width=1.0\linewidth]{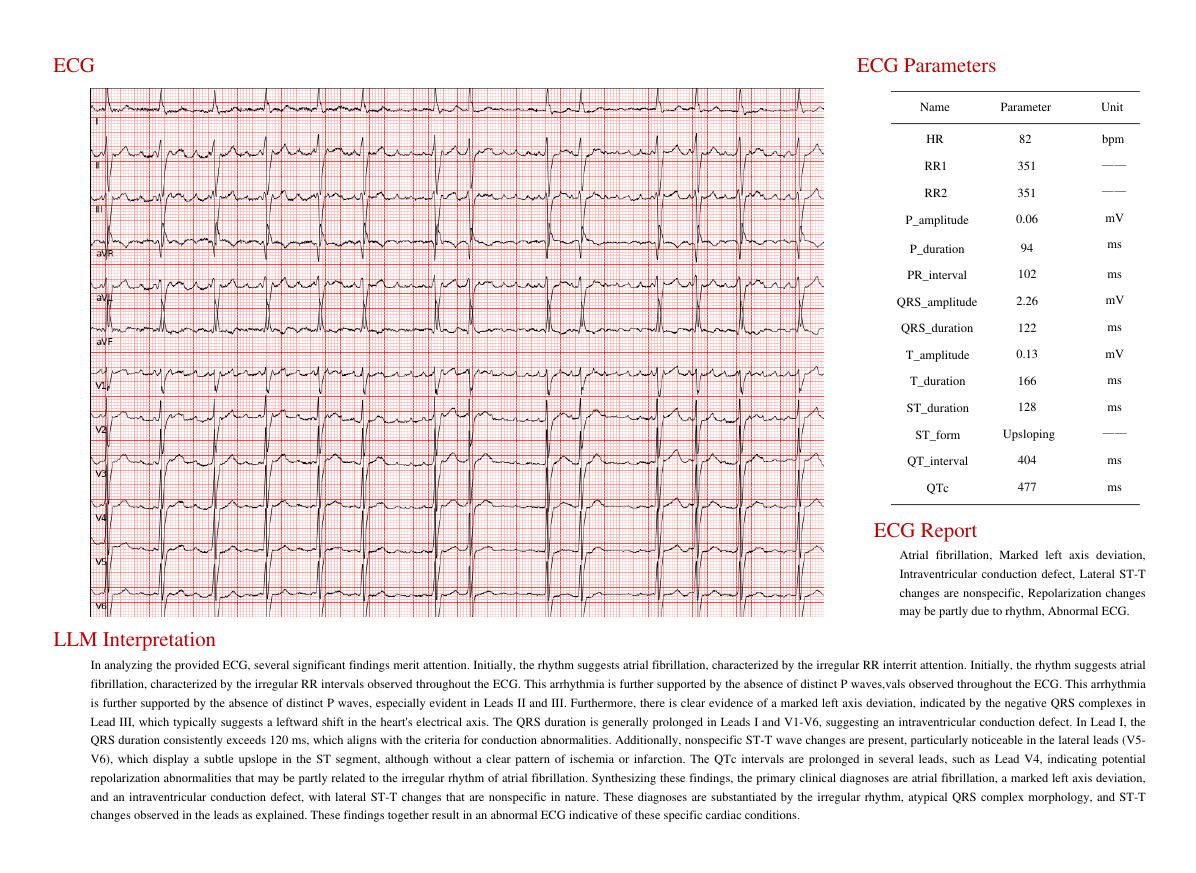}
 \caption{Multimodal ECG data.}
 \label{fig2}
\end{figure}

\section*{Usage Notes}

This module gives MEETI users an additional, potentially valuable source of information for research involving the ECG LLM. The dataset does have some limitations. Because of image size constraints, it includes about 10,000 ECG images. Researchers who need more ECG image data can use MIMIC-IV-ECG as a foundation to generate additional samples from the GitHub repository (\url{https://github.com/PKUDigitalHealth/ecg_plot}). Our recent work GEM \cite{lan2025gem} demonstrates how to leverage such multimodal ECG data to train MLLMs for ECG understanding.


\section*{RESOURCE AVAILABILITY}

The code of our method and evaluations is publicly available at \url{https://github.com/PKUDigitalHealth/ecg_plot} and \url{https://github.com/PKUDigitalHealth/FeatureDB}. All the datasets used in our work are publicly available: MIMIC-IV-ECG (\url{https://physionet.org/content/mimic-iv-ecg/1.0/}) and MEETI (\url{https://github.com/PKUDigitalHealth/MIMIC-IV-ECG-Ext-Text-Image}).

\section*{ACKNOWLEDGMENTS}

This work was supported by the CCF-Zhipu Large Model Innovation Fund (CCF-Zhipu202414); the Joint Fund for Medical Artificial Intelligence (MAI2022C004); the Fan Sumei scientific research start-up funds (DT2400000509).

\section*{AUTHOR CONTRIBUTIONS}


Creation and maintenance of the original database: D.Z., X.L., M.F. and S.H.; Conception of the release process: D.Z., X.L. and S.H.; Data harmonization: D.Z., X.L., S.G. and Q.Z.; Providing conversion routines: D.Z.; Manuscript preparation: D.Z., X.L., S.G., S.F. and S.H.; Supervision of the project: M.F. and S.H.; Critical comments and revision of manuscript: all authors.

\section*{DECLARATION OF INTERESTS}

The authors declare no competing interests.


\newpage


\bibliography{reference}

\begin{thebibliography}{28}
\providecommand{\natexlab}[1]{#1}
\providecommand{\url}[1]{\texttt{#1}}
\providecommand{\href}[2]{#2}
\providecommand{\path}[1]{#1}
\providecommand{\DOIprefix}{doi: }
\providecommand{\ArXivprefix}{arXiv: }
\providecommand{\URLprefix}{URL: }
\providecommand{\Pubmedprefix}{pmid: }
\providecommand{\doi}[1]{\href{http://dx.doi.org/#1}{\path{#1}}}
\providecommand{\Pubmed}[1]{\href{pmid:#1}{\path{#1}}}
\providecommand{\BIBand}{and}
\providecommand{\bibinfo}[2]{#2}
\ifx\xfnm\undefined \def\xfnm[#1]{\unskip,\space#1}\fi
\makeatletter\def\@biblabel#1{#1.}\makeatother
\bibitem[{Martin et~al.(2025)Martin, Aday, Allen, Almarzooq, Anderson, Arora, Avery, Baker-Smith, Bansal, Beaton et~al.}]{martin20252025}
\bibinfo{author}{Martin, S.S.}, \bibinfo{author}{Aday, A.W.}, \bibinfo{author}{Allen, N.B.}, \bibinfo{author}{Almarzooq, Z.I.}, \bibinfo{author}{Anderson, C.A.}, \bibinfo{author}{Arora, P.}, \bibinfo{author}{Avery, C.L.}, \bibinfo{author}{Baker-Smith, C.M.}, \bibinfo{author}{Bansal, N.}, \bibinfo{author}{Beaton, A.Z.} et~al. (\bibinfo{year}{2025}). \bibinfo{title}{2025 heart disease and stroke statistics: a report of us and global data from the american heart association}.
\newblock \bibinfo{journal}{Circulation} \emph{\bibinfo{volume}{151}}, \bibinfo{pages}{e41--e660}.
\bibitem[{Tan et~al.(2025)Tan, Fan, Luo, Zhou, Wang, Wang, Liu, Liu and Wang}]{tan2025pediatric}
\bibinfo{author}{Tan, J.}, \bibinfo{author}{Fan, H.}, \bibinfo{author}{Luo, J.}, \bibinfo{author}{Zhou, Y.}, \bibinfo{author}{Wang, N.}, \bibinfo{author}{Wang, X.}, \bibinfo{author}{Liu, G.}, \bibinfo{author}{Liu, C.}, and \bibinfo{author}{Wang, Z.} (\bibinfo{year}{2025}). \bibinfo{title}{A pediatric ecg database with disease diagnosis covering 11643 children}.
\newblock \bibinfo{journal}{Scientific Data} \emph{\bibinfo{volume}{12}}, \bibinfo{pages}{867}.
\bibitem[{Yagi et~al.(2024)Yagi, Mori, Goto, Iwami and Inoue}]{yagi2024routine}
\bibinfo{author}{Yagi, R.}, \bibinfo{author}{Mori, Y.}, \bibinfo{author}{Goto, S.}, \bibinfo{author}{Iwami, T.}, and \bibinfo{author}{Inoue, K.} (\bibinfo{year}{2024}). \bibinfo{title}{Routine electrocardiogram screening and cardiovascular disease events in adults}.
\newblock \bibinfo{journal}{JAMA Internal Medicine} \emph{\bibinfo{volume}{184}}, \bibinfo{pages}{1035--1044}.
\bibitem[{Zheng et~al.(2020{\natexlab{a}})Zheng, Fu, Anderson, Chu and Rakovski}]{zheng202012}
\bibinfo{author}{Zheng, J.}, \bibinfo{author}{Fu, G.}, \bibinfo{author}{Anderson, K.}, \bibinfo{author}{Chu, H.}, and \bibinfo{author}{Rakovski, C.} (\bibinfo{year}{2020}{\natexlab{a}}). \bibinfo{title}{A 12-lead ecg database to identify origins of idiopathic ventricular arrhythmia containing 334 patients}.
\newblock \bibinfo{journal}{Scientific data} \emph{\bibinfo{volume}{7}}, \bibinfo{pages}{98}.
\bibitem[{Hong et~al.(2020{\natexlab{a}})Hong, Zhou, Shang, Xiao and Sun}]{hong2020opportunities}
\bibinfo{author}{Hong, S.}, \bibinfo{author}{Zhou, Y.}, \bibinfo{author}{Shang, J.}, \bibinfo{author}{Xiao, C.}, and \bibinfo{author}{Sun, J.} (\bibinfo{year}{2020}{\natexlab{a}}). \bibinfo{title}{Opportunities and challenges of deep learning methods for electrocardiogram data: A systematic review}.
\newblock \bibinfo{journal}{Computers in biology and medicine} \emph{\bibinfo{volume}{122}}, \bibinfo{pages}{103801}.
\bibitem[{Lan et~al.(2022)Lan, Ng, Hong and Feng}]{lan2022intra}
\bibinfo{author}{Lan, X.}, \bibinfo{author}{Ng, D.}, \bibinfo{author}{Hong, S.}, and \bibinfo{author}{Feng, M.} (\bibinfo{year}{2022}). \bibinfo{title}{Intra-inter subject self-supervised learning for multivariate cardiac signals}.
\newblock In \bibinfo{booktitle}{Proceedings of the AAAI Conference on Artificial Intelligence} vol.~\bibinfo{volume}{36}. pp. \bibinfo{pages}{4532--4540}.
\bibitem[{Lan et~al.(2024)Lan, Yan, Hong and Feng}]{lan2024towards}
\bibinfo{author}{Lan, X.}, \bibinfo{author}{Yan, H.}, \bibinfo{author}{Hong, S.}, and \bibinfo{author}{Feng, M.} (\bibinfo{year}{2024}). \bibinfo{title}{Towards enhancing time series contrastive learning: A dynamic bad pair mining approach}.
\newblock In \bibinfo{booktitle}{The Twelfth International Conference on Learning Representations}.
\newblock \URLprefix \url{https://openreview.net/forum?id=K2c04ulKXn}.
\bibitem[{Li et~al.(2025)Li, Aguirre, Junior, Jin, Liu, Zhong, Sun, Clifford, Brandon~Westover and Hong}]{li2025electrocardiogram}
\bibinfo{author}{Li, J.}, \bibinfo{author}{Aguirre, A.D.}, \bibinfo{author}{Junior, V.M.}, \bibinfo{author}{Jin, J.}, \bibinfo{author}{Liu, C.}, \bibinfo{author}{Zhong, L.}, \bibinfo{author}{Sun, C.}, \bibinfo{author}{Clifford, G.}, \bibinfo{author}{Brandon~Westover, M.}, and \bibinfo{author}{Hong, S.} (\bibinfo{year}{2025}). \bibinfo{title}{An electrocardiogram foundation model built on over 10 million recordings}.
\newblock \bibinfo{journal}{NEJM AI} \emph{\bibinfo{volume}{2}}, \bibinfo{pages}{AIoa2401033}.
\bibitem[{Liu et~al.(2023)Liu, Zhu, Wu, Yang, You, Wang, Lu, Liu, Zheng, Sun et~al.}]{liu2023medical}
\bibinfo{author}{Liu, F.}, \bibinfo{author}{Zhu, T.}, \bibinfo{author}{Wu, X.}, \bibinfo{author}{Yang, B.}, \bibinfo{author}{You, C.}, \bibinfo{author}{Wang, C.}, \bibinfo{author}{Lu, L.}, \bibinfo{author}{Liu, Z.}, \bibinfo{author}{Zheng, Y.}, \bibinfo{author}{Sun, X.} et~al. (\bibinfo{year}{2023}). \bibinfo{title}{A medical multimodal large language model for future pandemics}.
\newblock \bibinfo{journal}{NPJ Digital Medicine} \emph{\bibinfo{volume}{6}}, \bibinfo{pages}{226}.
\bibitem[{Sandmann et~al.(2025)Sandmann, Hegselmann, Fujarski, Bickmann, Wild, Eils and Varghese}]{sandmann2025benchmark}
\bibinfo{author}{Sandmann, S.}, \bibinfo{author}{Hegselmann, S.}, \bibinfo{author}{Fujarski, M.}, \bibinfo{author}{Bickmann, L.}, \bibinfo{author}{Wild, B.}, \bibinfo{author}{Eils, R.}, and \bibinfo{author}{Varghese, J.} (\bibinfo{year}{2025}). \bibinfo{title}{Benchmark evaluation of deepseek large language models in clinical decision-making}.
\newblock \bibinfo{journal}{Nature Medicine} pp. \bibinfo{pages}{1--1}.
\bibitem[{Zhou et~al.(2025)Zhou, Xu, Zhang, Xu, Guo, Zhan, Fang, Ding, Wang, Xu et~al.}]{zhou2025large}
\bibinfo{author}{Zhou, S.}, \bibinfo{author}{Xu, Z.}, \bibinfo{author}{Zhang, M.}, \bibinfo{author}{Xu, C.}, \bibinfo{author}{Guo, Y.}, \bibinfo{author}{Zhan, Z.}, \bibinfo{author}{Fang, Y.}, \bibinfo{author}{Ding, S.}, \bibinfo{author}{Wang, J.}, \bibinfo{author}{Xu, K.} et~al. (\bibinfo{year}{2025}). \bibinfo{title}{Large language models for disease diagnosis: A scoping review}.
\newblock \bibinfo{journal}{npj Artificial Intelligence} \emph{\bibinfo{volume}{1}}, \bibinfo{pages}{9}.
\bibitem[{Hong et~al.(2025)Hong, Liu, Wu, Lu, Yang, Chen, Rao, Liu, Ye, Zhuang et~al.}]{hong2025evaluating}
\bibinfo{author}{Hong, Q.}, \bibinfo{author}{Liu, S.}, \bibinfo{author}{Wu, L.}, \bibinfo{author}{Lu, Q.}, \bibinfo{author}{Yang, P.}, \bibinfo{author}{Chen, D.}, \bibinfo{author}{Rao, G.}, \bibinfo{author}{Liu, X.}, \bibinfo{author}{Ye, H.}, \bibinfo{author}{Zhuang, P.} et~al. (\bibinfo{year}{2025}). \bibinfo{title}{Evaluating the performance of large language \& visual-language models in cervical cytology screening}.
\newblock \bibinfo{journal}{NPJ Precision Oncology} \emph{\bibinfo{volume}{9}}, \bibinfo{pages}{153}.
\bibitem[{Wagner et~al.(2020)Wagner, Strodthoff, Bousseljot, Kreiseler, Lunze, Samek and Schaeffter}]{wagner2020ptb}
\bibinfo{author}{Wagner, P.}, \bibinfo{author}{Strodthoff, N.}, \bibinfo{author}{Bousseljot, R.D.}, \bibinfo{author}{Kreiseler, D.}, \bibinfo{author}{Lunze, F.I.}, \bibinfo{author}{Samek, W.}, and \bibinfo{author}{Schaeffter, T.} (\bibinfo{year}{2020}). \bibinfo{title}{Ptb-xl, a large publicly available electrocardiography dataset}.
\newblock \bibinfo{journal}{Scientific data} \emph{\bibinfo{volume}{7}}, \bibinfo{pages}{1--15}.
\bibitem[{Gow et~al.(2023)Gow, Pollard, Nathanson, Johnson, Moody, Fernandes, Greenbaum, Waks, Eslami, Carbonati et~al.}]{gow2023mimic}
\bibinfo{author}{Gow, B.}, \bibinfo{author}{Pollard, T.}, \bibinfo{author}{Nathanson, L.A.}, \bibinfo{author}{Johnson, A.}, \bibinfo{author}{Moody, B.}, \bibinfo{author}{Fernandes, C.}, \bibinfo{author}{Greenbaum, N.}, \bibinfo{author}{Waks, J.W.}, \bibinfo{author}{Eslami, P.}, \bibinfo{author}{Carbonati, T.} et~al. (\bibinfo{year}{2023}). \bibinfo{title}{Mimic-iv-ecg: Diagnostic electrocardiogram matched subset}.
\newblock \bibinfo{journal}{Type: dataset} \emph{\bibinfo{volume}{6}}, \bibinfo{pages}{13--14}.
\bibitem[{Kansal et~al.(2025)Kansal, Chen, Jin, Rajpurkar and Kim}]{kansal2025mc}
\bibinfo{author}{Kansal, A.}, \bibinfo{author}{Chen, E.}, \bibinfo{author}{Jin, B.T.}, \bibinfo{author}{Rajpurkar, P.}, and \bibinfo{author}{Kim, D.A.} (\bibinfo{year}{2025}). \bibinfo{title}{Mc-med, multimodal clinical monitoring in the emergency department}.
\newblock \bibinfo{journal}{Scientific Data} \emph{\bibinfo{volume}{12}}, \bibinfo{pages}{1094}.
\bibitem[{Al-Zaiti et~al.(2023)Al-Zaiti, Martin-Gill, Z{\`e}gre-Hemsey, Bouzid, Faramand, Alrawashdeh, Gregg, Helman, Riek, Kraevsky-Phillips et~al.}]{al2023machine}
\bibinfo{author}{Al-Zaiti, S.S.}, \bibinfo{author}{Martin-Gill, C.}, \bibinfo{author}{Z{\`e}gre-Hemsey, J.K.}, \bibinfo{author}{Bouzid, Z.}, \bibinfo{author}{Faramand, Z.}, \bibinfo{author}{Alrawashdeh, M.O.}, \bibinfo{author}{Gregg, R.E.}, \bibinfo{author}{Helman, S.}, \bibinfo{author}{Riek, N.T.}, \bibinfo{author}{Kraevsky-Phillips, K.} et~al. (\bibinfo{year}{2023}). \bibinfo{title}{Machine learning for ecg diagnosis and risk stratification of occlusion myocardial infarction}.
\newblock \bibinfo{journal}{Nature Medicine} \emph{\bibinfo{volume}{29}}, \bibinfo{pages}{1804--1813}.
\bibitem[{Chen et~al.(2024)Chen, Huang, Zhang, Chang, Zhang, Li, Qiu, Hu, Peng, Du et~al.}]{chen2024congenital}
\bibinfo{author}{Chen, J.}, \bibinfo{author}{Huang, S.}, \bibinfo{author}{Zhang, Y.}, \bibinfo{author}{Chang, Q.}, \bibinfo{author}{Zhang, Y.}, \bibinfo{author}{Li, D.}, \bibinfo{author}{Qiu, J.}, \bibinfo{author}{Hu, L.}, \bibinfo{author}{Peng, X.}, \bibinfo{author}{Du, Y.} et~al. (\bibinfo{year}{2024}). \bibinfo{title}{Congenital heart disease detection by pediatric electrocardiogram based deep learning integrated with human concepts}.
\newblock \bibinfo{journal}{Nature Communications} \emph{\bibinfo{volume}{15}}, \bibinfo{pages}{976}.
\bibitem[{Fan et~al.(2025)Fan, Zhang, Wang, Geng, Lu, Sang, Xu, Wang, Zhao, Cheng et~al.}]{fan2025detecting}
\bibinfo{author}{Fan, S.}, \bibinfo{author}{Zhang, D.}, \bibinfo{author}{Wang, Y.}, \bibinfo{author}{Geng, S.}, \bibinfo{author}{Lu, K.}, \bibinfo{author}{Sang, M.}, \bibinfo{author}{Xu, W.}, \bibinfo{author}{Wang, H.}, \bibinfo{author}{Zhao, Q.}, \bibinfo{author}{Cheng, C.} et~al. (\bibinfo{year}{2025}). \bibinfo{title}{Detecting long qt syndrome and first-degree atrioventricular block using single-lead ai-ecg: A multi-center real-world study}.
\newblock \bibinfo{journal}{arXiv preprint arXiv:2502.17499}.
\bibitem[{{\c{S}}AHiN et~al.(2025){\c{S}}AHiN, Arslan and {\"O}zdemir}]{csahin2025unlocking}
\bibinfo{author}{{\c{S}}AHiN, E.}, \bibinfo{author}{Arslan, N.N.}, and \bibinfo{author}{{\"O}zdemir, D.} (\bibinfo{year}{2025}). \bibinfo{title}{Unlocking the black box: an in-depth review on interpretability, explainability, and reliability in deep learning}.
\newblock \bibinfo{journal}{Neural Computing and Applications} \emph{\bibinfo{volume}{37}}, \bibinfo{pages}{859--965}.
\bibitem[{Li et~al.(2024)Li, Liu, Cheng, Arcucci and Hong}]{li2024frozen}
\bibinfo{author}{Li, J.}, \bibinfo{author}{Liu, C.}, \bibinfo{author}{Cheng, S.}, \bibinfo{author}{Arcucci, R.}, and \bibinfo{author}{Hong, S.} (\bibinfo{year}{2024}). \bibinfo{title}{Frozen language model helps ecg zero-shot learning}.
\newblock In \bibinfo{booktitle}{Medical Imaging with Deep Learning}. \bibinfo{organization}{PMLR} pp. \bibinfo{pages}{402--415}.
\bibitem[{Lai et~al.(2025)Lai, Chen, Zhao, Zhang, Wang, Geng, Li and Hong}]{lai2025diffusets}
\bibinfo{author}{Lai, Y.}, \bibinfo{author}{Chen, J.}, \bibinfo{author}{Zhao, Q.}, \bibinfo{author}{Zhang, D.}, \bibinfo{author}{Wang, Y.}, \bibinfo{author}{Geng, S.}, \bibinfo{author}{Li, H.}, and \bibinfo{author}{Hong, S.} (\bibinfo{year}{2025}). \bibinfo{title}{Diffusets: 12-lead ecg generation conditioned on clinical text reports and patient-specific information}.
\newblock \bibinfo{journal}{Patterns}.
\bibitem[{Tsutsui et~al.(2025)Tsutsui, Brimer, Ben-Moshe, Sellal, Oster, Mori, Ikeda, Arai, Nakano, Kato et~al.}]{tsutsui2025shdb}
\bibinfo{author}{Tsutsui, K.}, \bibinfo{author}{Brimer, S.B.}, \bibinfo{author}{Ben-Moshe, N.}, \bibinfo{author}{Sellal, J.M.}, \bibinfo{author}{Oster, J.}, \bibinfo{author}{Mori, H.}, \bibinfo{author}{Ikeda, Y.}, \bibinfo{author}{Arai, T.}, \bibinfo{author}{Nakano, S.}, \bibinfo{author}{Kato, R.} et~al. (\bibinfo{year}{2025}). \bibinfo{title}{Shdb-af: a japanese holter ecg database of atrial fibrillation}.
\newblock \bibinfo{journal}{Scientific Data} \emph{\bibinfo{volume}{12}}, \bibinfo{pages}{454}.
\bibitem[{Zheng et~al.(2020{\natexlab{b}})Zheng, Zhang, Danioko, Yao, Guo and Rakovski}]{zheng2020121}
\bibinfo{author}{Zheng, J.}, \bibinfo{author}{Zhang, J.}, \bibinfo{author}{Danioko, S.}, \bibinfo{author}{Yao, H.}, \bibinfo{author}{Guo, H.}, and \bibinfo{author}{Rakovski, C.} (\bibinfo{year}{2020}{\natexlab{b}}). \bibinfo{title}{A 12-lead electrocardiogram database for arrhythmia research covering more than 10,000 patients}.
\newblock \bibinfo{journal}{Scientific data} \emph{\bibinfo{volume}{7}}, \bibinfo{pages}{48}.
\bibitem[{Johnson et~al.(2020)Johnson, Bulgarelli, Pollard, Horng, Celi and Mark}]{johnson2020mimic}
\bibinfo{author}{Johnson, A.}, \bibinfo{author}{Bulgarelli, L.}, \bibinfo{author}{Pollard, T.}, \bibinfo{author}{Horng, S.}, \bibinfo{author}{Celi, L.A.}, and \bibinfo{author}{Mark, R.} (\bibinfo{year}{2020}). \bibinfo{title}{Mimic-iv}.
\newblock \bibinfo{journal}{PhysioNet. Available online at: https://physionet. org/content/mimiciv/1.0/(accessed August 23, 2021)} pp. \bibinfo{pages}{49--55}.
\bibitem[{Hong et~al.(2020{\natexlab{b}})Hong, Fu, Zhou, Yu, Li, Wang and Cheng}]{hong2020cardiolearn}
\bibinfo{author}{Hong, S.}, \bibinfo{author}{Fu, Z.}, \bibinfo{author}{Zhou, R.}, \bibinfo{author}{Yu, J.}, \bibinfo{author}{Li, Y.}, \bibinfo{author}{Wang, K.}, and \bibinfo{author}{Cheng, G.} (\bibinfo{year}{2020}{\natexlab{b}}). \bibinfo{title}{Cardiolearn: a cloud deep learning service for cardiac disease detection from electrocardiogram}.
\newblock In \bibinfo{booktitle}{Companion Proceedings of the Web Conference 2020}. pp. \bibinfo{pages}{148--152}.
\bibitem[{Zhang et~al.(2023)Zhang, Geng, Zhou, Xu, Wei, Wang, Yu, Zhu, Li, Zhao et~al.}]{zhang2023artificial}
\bibinfo{author}{Zhang, D.}, \bibinfo{author}{Geng, S.}, \bibinfo{author}{Zhou, Y.}, \bibinfo{author}{Xu, W.}, \bibinfo{author}{Wei, G.}, \bibinfo{author}{Wang, K.}, \bibinfo{author}{Yu, J.}, \bibinfo{author}{Zhu, Q.}, \bibinfo{author}{Li, Y.}, \bibinfo{author}{Zhao, Y.} et~al. (\bibinfo{year}{2023}). \bibinfo{title}{Artificial intelligence system for detection and screening of cardiac abnormalities using electrocardiogram images}.
\newblock \bibinfo{journal}{arXiv preprint arXiv:2302.10301}.
\bibitem[{Achiam et~al.(2023)Achiam, Adler, Agarwal, Ahmad, Akkaya, Aleman, Almeida, Altenschmidt, Altman, Anadkat et~al.}]{achiam2023gpt}
\bibinfo{author}{Achiam, J.}, \bibinfo{author}{Adler, S.}, \bibinfo{author}{Agarwal, S.}, \bibinfo{author}{Ahmad, L.}, \bibinfo{author}{Akkaya, I.}, \bibinfo{author}{Aleman, F.L.}, \bibinfo{author}{Almeida, D.}, \bibinfo{author}{Altenschmidt, J.}, \bibinfo{author}{Altman, S.}, \bibinfo{author}{Anadkat, S.} et~al. (\bibinfo{year}{2023}). \bibinfo{title}{Gpt-4 technical report}.
\newblock \bibinfo{journal}{arXiv preprint arXiv:2303.08774}.
\bibitem[{Lan et~al.(2025)Lan, Wu, He, Zhao, Hong and Feng}]{lan2025gem}
\bibinfo{author}{Lan, X.}, \bibinfo{author}{Wu, F.}, \bibinfo{author}{He, K.}, \bibinfo{author}{Zhao, Q.}, \bibinfo{author}{Hong, S.}, and \bibinfo{author}{Feng, M.} (\bibinfo{year}{2025}). \bibinfo{title}{Gem: Empowering mllm for grounded ecg understanding with time series and images}.
\newblock \bibinfo{journal}{arXiv preprint arXiv:2503.06073}.

\end{thebibliography}

\bigskip


\newpage

\end{document}